\documentclass[useAMS,usenatbib]{mn2e}

\usepackage{fancyhdr}
\usepackage{hyperref}
\usepackage{mn2e-bst}
\usepackage{graphicx}
\usepackage{CJKutf8}
\usepackage{amssymb,amsmath}
\newif\ifAMStwofonts
\AMStwofontstrue
\newcommand{\aj}{AJ}			% Astronomical Journal
\newcommand{\araa}{ARA\&A}		% Annual Review of Astron and Astrophys
\newcommand{\apj}{ApJ}			% Astrophysical Journal
\newcommand{\apjl}{ApJLett}		% Astrophysical Journal, Letters
\newcommand{\apjs}{ApJS}		% Astrophysical Journal, Supplement
\newcommand{\aap}{A\&A}			% Astronomy and Astrophysics
\newcommand{\mnras}{MNRAS}		% Monthly Notices of the RAS
\newcommand{\gca}{Geochim.~Cosmochim.~Acta}   % Geochimica Cosmochimica Acta
\newcommand{\jcap}{JCAP}

\def\ltsima{$\; \buildrel < \over \sim \;$}
\def\simlt{\lower.5ex\hbox{\ltsima}}
\def\gtsima{$\; \buildrel > \over \sim \;$}
\def\simgt{\lower.5ex\hbox{\gtsima}}

  \newcommand{\bc}{\begin{center}}
  \newcommand{\ec}{\end{center}}
  
  \newcommand{\hMsun}{~h^{-1}\>{\rm M_\odot}}
  
  \newcommand{\Kpc}{~h^{-1}~{\rm kpc}}
  \newcommand{\Hunit}{~h ~{\rm km}~s^{-1}~{\rm Mpc}^{-1}}

  \title[On the Estimation of Galaxy Clusters Centres]{How does our choice of
    observable influence our estimation of the centre of a galaxy cluster?
    Insights from cosmological simulations.}

\author[Weiguang Cui, et al.]
{\parbox{\textwidth}{Weiguang Cui,$^{1,2}$\thanks{E-mail: \texttt{weiguang.cui@uwa.edu.au}}
  Chris Power$^{1,2}$, Veronica Biffi${^3}$, Stefano Borgani$^{3,4,5}$, Alexander Knebe$^{6,7}$,
  Giuseppe Murante$^3$, Dunja Fabjan$^{8,4}$, Geraint F. Lewis$^9$ \&
  Greg B. Poole$^{10}$ }\vspace{0.4cm}\\
\parbox{\textwidth}{$^1$ ICRAR, University of Western Australia, 35 Stirling Highway,
Crawley, Western Australia 6009, Australia\\
$^2$ ARC Centre of Excellence for All-Sky Astrophysics (CAASTRO)\\
$^3$ Astronomy Unit, Department of Physics, University of Trieste,
via Tiepolo 11, I-34131 Trieste, Italy\\
$^4$ INAF -- Astronomical Observatory of Trieste, via Tiepolo 11,
I-34131 Trieste, Italy\\
$^5$ INFN -- Sezione di Trieste, I-34100 Trieste, Italy\\
$^6$Departamento de F\'isica Te\'{o}rica, M\'{o}dulo 15, Facultad de Ciencias,
Universidad Aut\'{o}noma de Madrid, 28049 Madrid, Spain\\
$^7$Astro-UAM, UAM, Unidad Asociada CSIC\\
$^8$Faculty of Mathematics and Physics, University of Ljubljana, Jadranska ulica
19, SI-1000 Ljubljana, Slovenia\\
$^9$ Sydney Institute for Astronomy School of Physics, A28, The University of Sydney NSW
2006, Australia \\
$^{10}$ School of Physics, University of Melbourne, Parksville, VIC 3010, Australia
}}

\begin{document}

%\date{Accepted ???. Received ???; in original form }

\pagerange{\pageref{firstpage}--\pageref{lastpage}} \pubyear{2015}

\maketitle

\label{firstpage}
\begin{abstract}
  Galaxy clusters are an established and powerful test-bed for theories of both
  galaxy evolution and cosmology. Accurate interpretation of cluster
  observations often requires robust identification of the location of the
  centre. Using a statistical sample of clusters drawn from a suite of
  cosmological simulations in which we have explored a range of galaxy formation
  models, we investigate how the location of this centre is affected by the
  choice of observable -- stars, hot gas, or the full mass distribution as can
  be probed by the gravitational potential. We explore several measures of
  cluster centre: the minimum of the gravitational potential, which would
  expect to define the centre if the cluster is in dynamical equilibrium; the
  peak of the density; the centre of BCG; and the peak and
  centroid of X-ray luminosity. We find that the centre of BCG
  correlates more strongly with the minimum of the gravitational potential than
  the X-ray defined centres, while AGN feedback acts to significantly enhance
  the offset between the peak X-ray luminosity and minimum gravitational
  potential. These results highlight the importance of centre identification
  when interpreting clusters observations, in particular when comparing
  theoretical predictions and observational data.
\end{abstract}
\begin{keywords}
	cosmology: theory -- galaxies: clusters: general -- galaxies: formation
\end{keywords}

%*****************************************************************************

\section{Introduction}
\label{i}

Currently favoured models of cosmological structure formation are hierarchical -- lower
mass systems merge progressively to form more massive structures, with galaxy clusters
representing the final state of this process. They are widely used as cosmological probes
\citep[e.g][]{Linden2014,Mantz2015}, but they are also unique laboratories for
testing models of gravitational structure formation, galaxy evolution, thermodynamics of
the intergalactic medium, and plasma physics \citep[e.g.][]{KravtsovBorgani2012}.

Observationally, galaxy clusters are usually identified through optical images
\citep[e.g.][]{Postman1996, Gladders2000, Ramella2001, Koester2007, Robotham2011}, X-ray
observations \citep[e.g.][]{Ebeling1998, Bohringer2004, Liu2013}, the
Sunyaev-Zel'dovich effect \citep[e.g.][]{Vanderlinde2010, Planck2011, Williamson2011},
and weak and strong gravitational lensing \citep[e.g.][]{Johnston2007,
  Mandelbaum2008, Zitrin2012}.
A fundamental step in any of these procedures is identification of the cluster
centre. For example, it is natural to adopt the optical/X-ray luminosity
peak/centroid or brightest cluster galaxy (BCG) position as the centre of an
optically or X-ray selected cluster respectively, whereas the location of the
minimum of the lensing potential is more natural when considering strong and
weak lensing.

It is interesting to ask how observational estimates of the cluster centre
relate to assumptions about the underlying physical mass distribution. This
can have important consequences for our interpretation of observations,
potentially biasing recovery of properties such as mass and concentration
\citep[e.g.][]{Shan2010b,Du2014}.
Theoretically, it is natural to select the location of the minimum of the
gravitational potential as the cluster centre, provided the cluster is
dynamically relaxed. If the hot X-ray emitting intra-cluster gas is in
hydrostatic equilibrium within the cluster potential and orbiting
stars are in dynamical equilibrium, then we should expect good agreement between
these different observable centre tracers and the potential minimum. However,
typical clusters are not in dynamical equilibrium -- they form
relatively recently and have undergone or are undergoing significant merging
activity, resulting in disturbed mass distributions
\citep[e.g.][]{Thomas98, Power2012} -- and so we might anticipate
systematic offsets between optical, X-ray and potential centres.

The goal of this paper is to estimate the size of offset that we might
expect by using a statistical sample of simulated galaxy clusters to measure
cluster centres as determined by different observables (e.g. centre of BCG,
X-ray emitting hot gas) and the minimum of the gravitational potential. We also
assess how these measurements are affected by AGN feedback,
which we would expect to influence the distribution of hot gas, but could also
influence when and where stars form. Before we present the results of our
analysis, we review briefly results from observations.\\

We argued that typical clusters are not in dynamical equilibrium, and so we
should expect offsets
between centres estimated using different tracers. This is borne out by
observations, which suggest that where one locates a cluster's centre will
depend on the choice of the tracer. \cite{Lin2004} looked at the offsets between
BCGs and X-ray peaks (or centroids) and found that about 75 per cent of
identified clusters had offsets within $0.06\,r_{200}$ (where $r_{200}$ is the
radius within which the enclosed mean matter overdensity is 200 times the
critical density of the Universe), 90 per cent within $0.38\,r_{200}$, and $\sim
10$ per cent contamination level
of possibly misidentified BCGs. \cite{Mann2012} found that the offsets between
BCGs and X-ray peaks are well approximated by a log-normal distribution, centred
at $\sim 11.5$ kpc; the typical offset between BCGs and X-ray centroids is
slightly larger at $\sim 21$ kpc. \cite{Rozo2014} found that $\sim 80$ per cent
of their clusters have a perfect agreement ($\simlt 50 $ kpc) between the X-ray
centroid and the central galaxy position; interestingly, the remaining clusters
were undergoing ongoing mergers and had offsets $\lesssim 300 $ kpc \citep[see
also][for similar findings]{Linden2014}. \cite{Zitrin2012} found that the
offset between a BCG's location and the peak of smoothed dark matter density is
well described by a log-normal distribution centred around $\sim 12.7 \Kpc$, and
the size of this offset increases with redshift, while \cite{Shan2010a}
characterized the offsets between the X-ray peaks and strong lensing centres and
found that about 45 per cent of their clusters show offsets of order $40-200 \Kpc$.

Identifying cluster centres observationally is not straightforward, however. For example,
\cite{Oguri2010} found that the distribution of separations between the location of the BCG
and the lensing centre has a long tail, and that the typical error on the mass centroid
measurement in weak lensing is $\sim 50 \Kpc$. \cite{George2012} found BCGs are one of the
best tracers of a cluster's centre-of-mass, with offsets typically less than 75 $\Kpc$,
but these measurements are susceptible to how the centre is defined (e.g. intensity
centroids vs intensity peaks) and this can cause a $5-30$ per cent bias in stacked weak
lensing analyses. Also, evidence of recent or ongoing merging activity
correlates with increased offsets, as revealed by, for example, the
\cite{Rozo2014} result mentioned already. Interestingly, the centroid shift
(offsets of a system’s X-ray surface brightness peak from its centroid) is
usually a good indicator of a cluster's dynamical state and recent merging
activity \citep[e.g.][]{Mohr1993,Poole2006}. Large offsets between the centre of
mass and the minimum of the gravitational potential have been shown to be good
indicators of recent merging activity and systems that are out of dynamical
equilibrium \citep[e.g.][]{Thomas98,Power2012}.

We note briefly that measurements of velocity offsets in groups and clusters
also imply spatial offsets. For example, \cite{Bosch2005} estimated that central
galaxies oscillate about the potential minimum with an offset of $\sim 3$ per
cent of the virial radius, using the difference between the velocity of central
galaxy and the average velocity of the satellites. Following this work, \cite{Guo2015}
analyzed CMASS BOSS galaxies and found this offset translates to a mean
projected radius of $\sim 0.3$ per cent of $R_{vir}$, or $\sim 1 - 3 \Kpc$ at
halo mass of $\log M = 13 - 13.6$.

This brief survey of observational results makes clear that cluster centre
identification is non-trivial, susceptible to both observational and
astrophysical uncertainties. Indeed, the three commonly adopted centre
tracers -- BCGs, X-ray and lensing -- do not agree with each other with
offsets from tens to several hundreds kpc.

We use simulations to examine how the choice of tracer population (e.g. stellar
luminosity weighted vs X-ray emission weighted vs lensing centre) affects our estimates
of the centre. Here we have unambiguous information about clusters -- they are identified
using an automated method in (at least) 3D using methods such as Friends of Friends
\cite[FoF,][]{Davis1985} or Spherical Overdensity \cite[SO,][]{Lacey1994}, the
results of which are in broad agreement as established by comparison projects such as
\cite{Knebe2011} and \cite{Knebe2013}. Typically the location of the minimum of
the potential is identified with the halo centre in the FoF algorithm \citep[for
  example,][]{Springel2001, Gabriella2004, Dolag2009}, and with the
maximum density peak in the SO algorithm, which can be deduced iteratively
\citep[e.g.][]{Tinker2008}, using an adaptive mesh
\citep[e.g. \small{AHF},][]{Knollmann2009, Gill2004a}, or via an SPH-style
density evaluation \citep[e.g. \small{PIAO},][]{Cui2014b}.
In this paper, we use the SO method to identify halos and compute the location of the
density peak using an SPH kernel approach. If we can better understand the astrophysical
origin of observed centre offsets, then we can recover more accurate measurements of
cluster mass profiles \citep[e.g.][]{Shan2010b}, reconstruction of assembly histories
\citep[e.g.][]{Mann2012}, and tests of cosmological models with cases such as bullet
clusters \citep[e.g.][]{ForeroRomero2010}.

In the following sections, we describe how we have used cosmological
hydro-simulations with different baryon models \citep[see also ][]{Cui2012a,
  Cui2014b} to
select the statistical sample of clusters (\S\ref{simulation}), and
we describe our cluster centre identification methods (\S\ref{method}). In
section \ref{results} we present the results of our
analysis, showing how measured offsets depend on the choice of tracer
population, and on the assumed baryon models. Finally, we summarise
our results in \S\ref{concl}, and comment on their significance for interpretation of
observations of galaxy clusters.

\section{The Simulated Galaxy Cluster Catalogue}
\label{simulation}

%%%%%%%%%% FIG 1 %%%%%%%%
\begin{figure*}
\includegraphics[width=1.0\textwidth]{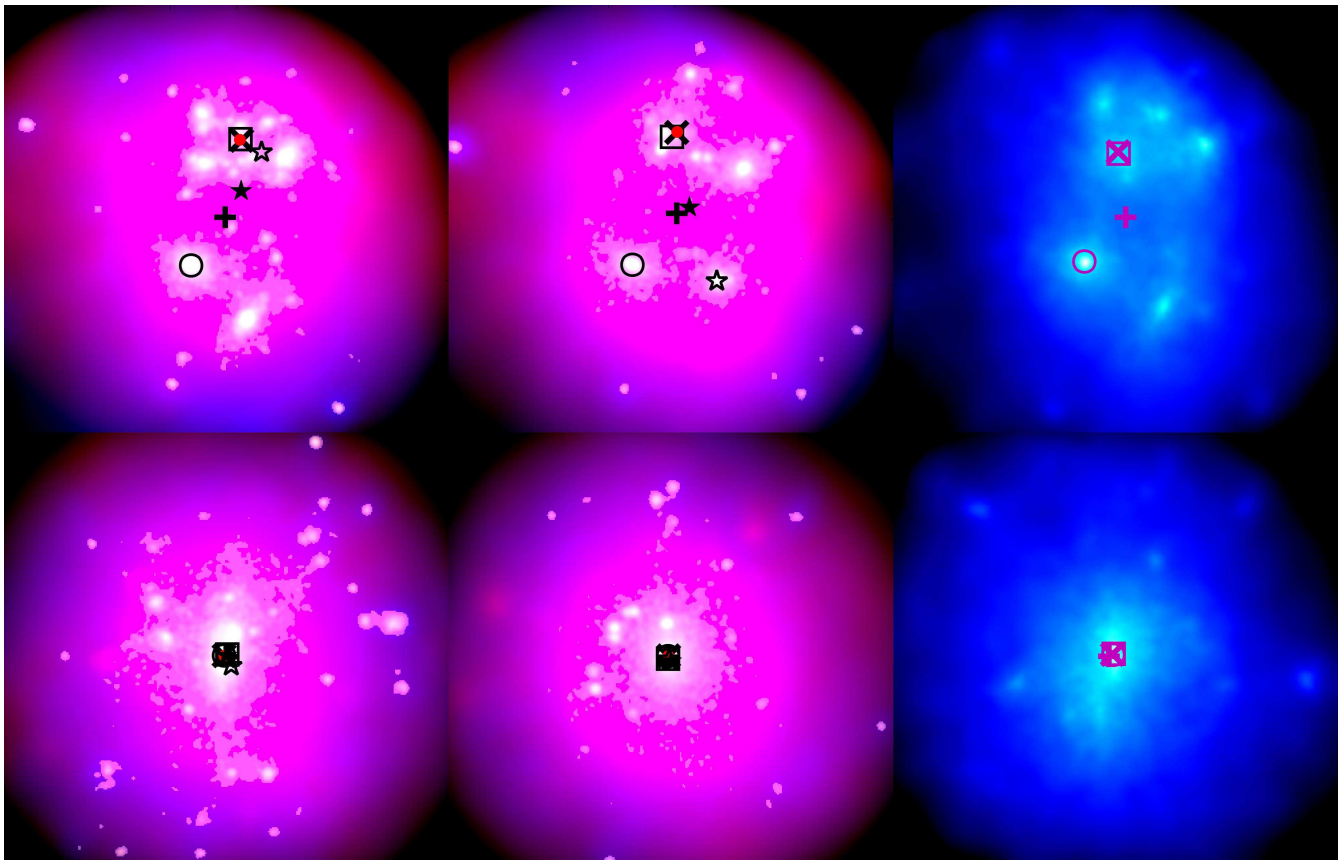}
\caption{Examples of one visually and dynamically disturbed (upper panel) and
  one undisturbed (lower panel) galaxy cluster at $z = 0$ from our suite of
  simulations (AGN, CSF, DM, from left to right). For the hydrodynamical
  simulated clusters, we use
  blue and red colours to represent dark matter and gas particle (SPH) densities,
  white represents optical stellar luminosity with a surface brightness of
  $\mu \geq 26.5 ~ mag/arcsec^2$ in the SDSS r band; the DM only equivalent
  is shown in the rightmost panel. The symbols (+,x,$\circ$,$\square$) identify
  the location of the cluster centre of mass (+); minimum of the gravitational
  potential (x); maximum of SPH kernel weighted density ($\circ$); and the
  iterative centre of mass ($\square$). For the two hydrodynamical runs, we show
  also the BCG position in SDSS $r$ band using red filled circles. The open and
  filled black star symbols indicate the X-ray peak
  and centroid positions, respectively. We refer to section \S\ref{method} for more
  details of these centre definitions.}
\label{fig:forshow}
\end{figure*}
%%%%%%%%%%%%%%%%%%%%%%%%%

We use three large--volume cosmological simulations, namely two hydrodynamical
simulations in which we include different feedback processes, and one dark matter
only N-body simulation. All these simulations are described in
\cite{Cui2012a, Cui2014b}; here we summarise the relevant details.

We assume a flat $\Lambda$CDM cosmology, with cosmological parameters of
$\Omega_{\rm m} = 0.24$ for the matter density parameter,
$\Omega_{\rm b} = 0.0413$ for the baryon contribution, $\sigma_8=0.8$
for the power spectrum normalisation, $n_{\rm s} = 0.96$ for the primordial
spectral index, and $h =0.73$ for the Hubble parameter in units of $100 \Hunit$.
The three simulations were set up using the same realisation of the
initial matter power spectrum, and reproduce the same large-scale structures.
We refer to the dark matter only simulation as the DM run. Both hydrodynamical
simulations include radiative cooling, star formation and kinetic feedback from
supernovae; in one case we ignore feedback from AGN (which is referred as the
CSF run), while in the other we include it (which is referred as the AGN run).

We use the TreePM-SPH code {\small GADGET-3}, an improved version of the public
{\small GADGET-2} code \citep{Gadget2}, which includes a range of prescriptions
for galaxy formation physics (e.g. cooling, star formation, feedback).
Gravitational forces are computed using a Plummer equivalent softening
fixed to $\epsilon_{Pl} = 7.5 \Kpc$ from z = 0 to 2 and fixed in comoving units
at a higher redshift. As we will see, our softening length $7.5 \Kpc$ is
comparable to -- and in cases larger than -- the offsets between the minimum
potential and maximum SPH density positions,
centre of BCGs and X-ray emission-weighted centres. However, the
minimum potential position is determined by the whole cluster, which should be
less affected by the softening length. Thus, we expect that these offsets are
accurate to within a softening length.

Haloes are identified using the spherical overdensity (SO) algorithm {\small PIAO}
\citep{Cui2014b}, assuming an overdensity criterion of $\Delta_c = 200$
\footnote{In the following, the overdensity value $\Delta_c$ is expressed in
  units of the cosmic critical density at a given redshift,
  $\rho_c(z)=3H^2(z)/(8\pi G)$.}. Densities are computed using a SPH kernel
smoothed over the nearest 128 neighbours; this allows us to determine the maximum
density in the halo, which we also identify as the density-weighted centre of the
halo. All of the particle types (dark matter, gas, stars) contribute equally
to the density computation.
% Since we focus on the cluster scale halos, which are unlikely to overlap, we allow the
% halos to overlap with each other in the code.

We select our cluster sample from the DM run SO halo catalogue, with the
requirement that $M_{200} > 2.0 \times 10^{14} \hMsun$; this gives a total of 184
halos in our sample, with a maximum mass of $\sim 1.2 \times 10^{15} \hMsun$.
The corresponding SO halos in the AGN and CSF runs are identified by cross-matching
the dark matter components using the unique particle IDs \cite[see more details
  in][]{Cui2014b}; we find no systems
less massive than $1.7 \times 10^{14} \hMsun$ in this cross-matched catalogue.
In this paper, we only focus on the clusters at redshift $z = 0$.

Examples of a visually and dynamically disturbed and undisturbed clusters (lower
and upper panels respectively) at $z = 0$ are shown in
Fig. \ref{fig:forshow}, where we show qualitative projected density distributions
in the AGN, CSF and DM runs (from left to right). In the case of the dark matter maps
(rightmost panels), only the dark matter contributes to the RGB value of a pixel. The
projected density of dark matter within a pixel lies in the range (0,255), and
this is used to set the ``B" of the RGB value of the pixel; if this
density exceeds a threshold, we set the RGB value to white. When combining dark
matter, gas and stars (leftmost and middle panels), both the dark matter and gas
contribute to the RGB value. As before, the projected density of dark matter is
scaled to the range (0,255), but without a threshold, and it is used to set the
``B" of the RGB value; the projected density of has is scaled to the
range (0,255) and is used to set the ``R" of the RGB value; and the RGB
value of stars is set to white, with a transparency of 0.5. By constructing the
projected density maps in this way, we can get a sense for the relative
projected densities of dark matter and gas in the systems; the projected dark
matter density dominates the hot gas density at larger radii in both systems,
but is dominated by the hot gas density at smaller radii.

\section{The cluster centre identification}
\label{method}

In this paper, we focus on 4 different definitions of the cluster centre.
We quote centres of potential and density, which are readily measured in
the simulation data, by their 3D values, while we use projected (2D) values
for centres derived from mock observational data.

\paragraph*{Minimum of the Gravitational Potential:} This is the physically
intuitive definition of the cluster centre, and is expected to correspond to the
lensing centre. For all particles within the $r_{200}$ radius, we select the one
with the most negative value of the potential as the cluster centre. The
particle's potential is directly coming from the simulations. We will take this
minimum potential position as the base line for comparison in this paper.

%% \paragraph*{Centre of Mass:} For all material within the $r_{200}$ radius, we
%% compute
%% \be
%% {\mathbf R}_c = \frac{1}{M} \sum\limits_{i=1}^n m_i {\mathbf r}_i;
%% \ee
%% here, $m_i$ is $i$th the particle mass, ${\mathbf r}_i$ is the position of
%% the particle, and $M$ is the total mass of the halo and n is the total number of particles
%% within $r_{200}$.

\paragraph*{Maximum of the SPH Density:} In constructing our halo catalogue using the
SO algorithm implemented in {\small PIAO}, we estimate the densities of particles
by smoothing over nearest neighbours using the SPH kernel, and identify the
particle with the highest density as the halo centre.\footnote{Although we employ this
  particular density estimate in this paper, we
  note that there are several methods to locate the centre when using the SO
  algorithm; in appendix \ref{A:denspeak}, we show how three different
  density peak estimators differ.}

\paragraph*{Optical Centres of the BCG:} Our hydrodynamical CSF and AGN runs
include star formation. Using the method applied in \cite{Cui2011}, we assign
luminosities to each of the star particles that form by assuming that they
constitute single stellar populations with ages, metallicities and masses given
the corresponding particle's properties in the run. Adopting the same initial
mass function as the simulation, the spectral energy distribution of each
particle is computed by interpolating the simple stellar population templates
of \cite{BC03}. We consider the three standard SDSS $r$, $g$, and $u$ bands in
this paper. The luminosity of each star particle is smoothed to a 2-D map
(projected to the xy-plane), with each pixel having a size of $5 \Kpc$. We
adopt the same spline kernel used for the SPH calculations with 49 SPH
neighbours, which is equivalent to $30 \Kpc$ \citep[see ][for more details]{Cui2014a}.
Note that the minimum offset cut for later
relevant plots will be set to half the image pixel size, $2.5 \Kpc$.

The centre of BCG is identified as the most luminous image
pixel of each band within the BCG. To select the BCG, we first separate the
intra-cluster light from galaxies. As shown in Fig. \ref{fig:forshow}, the surface
brightness cut ($\mu \geq 26.5 ~ mag/arcsec^2$) employed observationally is not
suitable for our simulated data because it would include too much intra-cluster
light.
\cite{Cui2014a} has shown that the physical intra-cluster light identification
method \citep[based on the star's velocity information][]{Dolag2010} implies
much higher surface brightness threshold values. For this reason, we adopt the
surface brightness threshold values, $\mu = 23, 24.75 ~ mag/arcsec^2$ for the
CSF and AGN runs, respectively. Although these two values are for V-band
luminosity in \cite{Cui2014a}, we apply them here to the three SDSS bands
without further corrections. This is because we are only interested in
position of the brightest pixel inside the BCG in this paper; corrections
should not affect our final results.
Pixels above the surface brightness threshold are grouped
together to form a galaxy by linking all neighbouring pixels,
starting from the brightest pixel. The most luminous galaxy is selected as the
BCG. In each band, we select the centre of the most luminous pixel
inside the BCG as the centre.

\paragraph*{Centres of X-ray Emission:} We estimate the X-ray emission from
each of the simulated clusters using the {\small PHOX} code
\citep[see][for a more detailed description]{Biffi2012, Biffi2013}. Specifically,
we simulate the X-ray emission of the intra-cluster medium (ICM) by adopting an
absorbed APEC model \cite[][]{Smith2001}, where the WABS absorption model
\cite[][]{Morrison1983} is used to mimic the Galactic absorption and the main
contribution from the hot ICM comes in the form of bremsstrahlung continuum plus
metal emission lines. The latter is obtained from the implementation of the APEC
model for a collisionally-ionized plasma comprised within the
XSPEC\footnote{http://heasarc.gsfc.nasa.gov/xanadu/xspec/.} package (v.12.8.0).
For any gas element in the simulation output, the model spectrum predicts the
expected number of photons, with which we statistically sample the spectral
energy distribution.

In the approach followed by PHOX, the synthetic X-ray photons are obtained from
the ideal emission spectrum calculated for every gas particle belonging to the
cluster ICM, depending on its density, temperature, metallicity
\footnote{In this work, a fiducial average metallicity of $Z =0.2\,Z_{\odot}$ is
  assumed, for simplicity, with solar abundances according to \citet{Angr1989}.}
and redshift (we assume $z = 0.05$ for the X-ray luminosity and angular-diameter
distances). We consider only the position of the X-ray centre in this
work, and do not expect the particular choice of redshift or metallicity to
affect it significantly. To
obtain the photon maps, we assume a realistic exposure time of
50\,ks and convolve the ideal photon-list of every cluster with the response
matrices of Chandra (ACIS-S detector); this accounts for the instrument
characteristics and sensitivity to the incoming photon energies. In this process,
the maps (i) are originally centered on the cluster potential centre, (ii)
cover a circular region of $R_{200}$ radius, and (iii) have a the same pixel size of
$5 \Kpc$ as the optical image.

In this work, we consider the $x$-$y$ projection and the full energy band of the
detector. In addition, we also apply the same SPH smoothing procedure as
used for the optical image, but using each pixel's photon counts from the
PHOX X-ray maps instead of stellar luminosity. The X-ray peak position is
identified as the pixel with the maximum value of photon counts. We note
here that using this simple X-ray peak position as the X-ray centre can be
biased by the satellites \citep[see][for more discussions about different X-ray centre
tracers]{Mantz2015}. The centroid of the X-ray map is computed basing on the method of
\cite{Bohringer2010, Rasia2013a}, modified to take the X-ray peak position as
the initial centre and reset to the centre of mass from photon counts within
the shrinking radius after each iteration. We reduce the radius to $85$ per
cent of the previous iteration, starting at an initial radius of $R_{200}$ in
projection, until a fixed inner radius $R_{2500}$ is reached. The X-ray centroid
is the centre of mass position at the final
step. We use this iterative method to locate the centroid, because there are
many un-relaxed clusters in our sample. Note that the minimum offset
cut for later relevant plots is also set to the size of half a pixel, $2.5 \Kpc$.

\section{Results}
\label{results}
%%%%%%%%%% FIG 2 %%%%%%%%
\begin{figure}
\centering
\begin{tabular}{@{}c@{}}
  \includegraphics[width=0.5\textwidth]{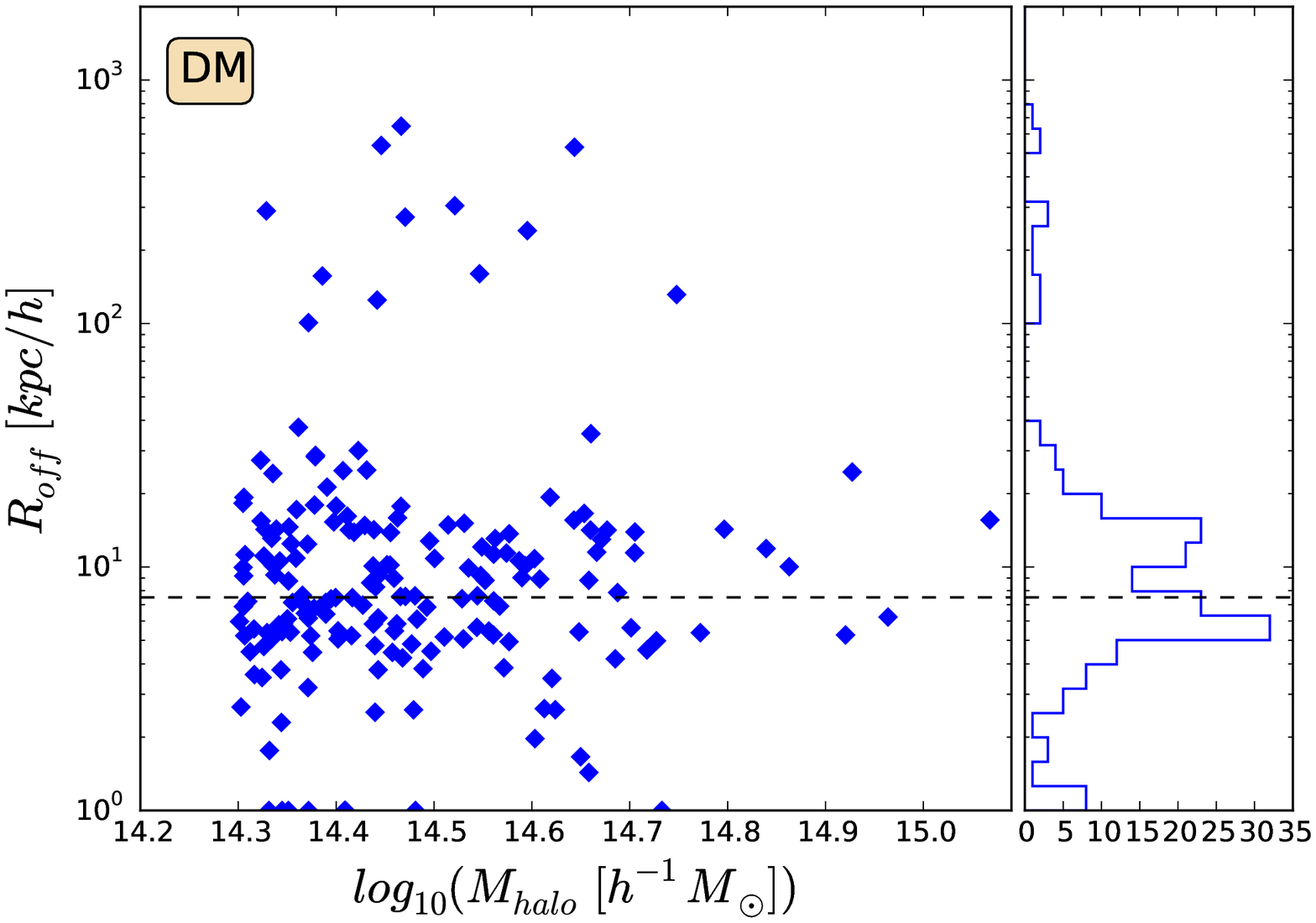} \\
  \includegraphics[width=0.5\textwidth]{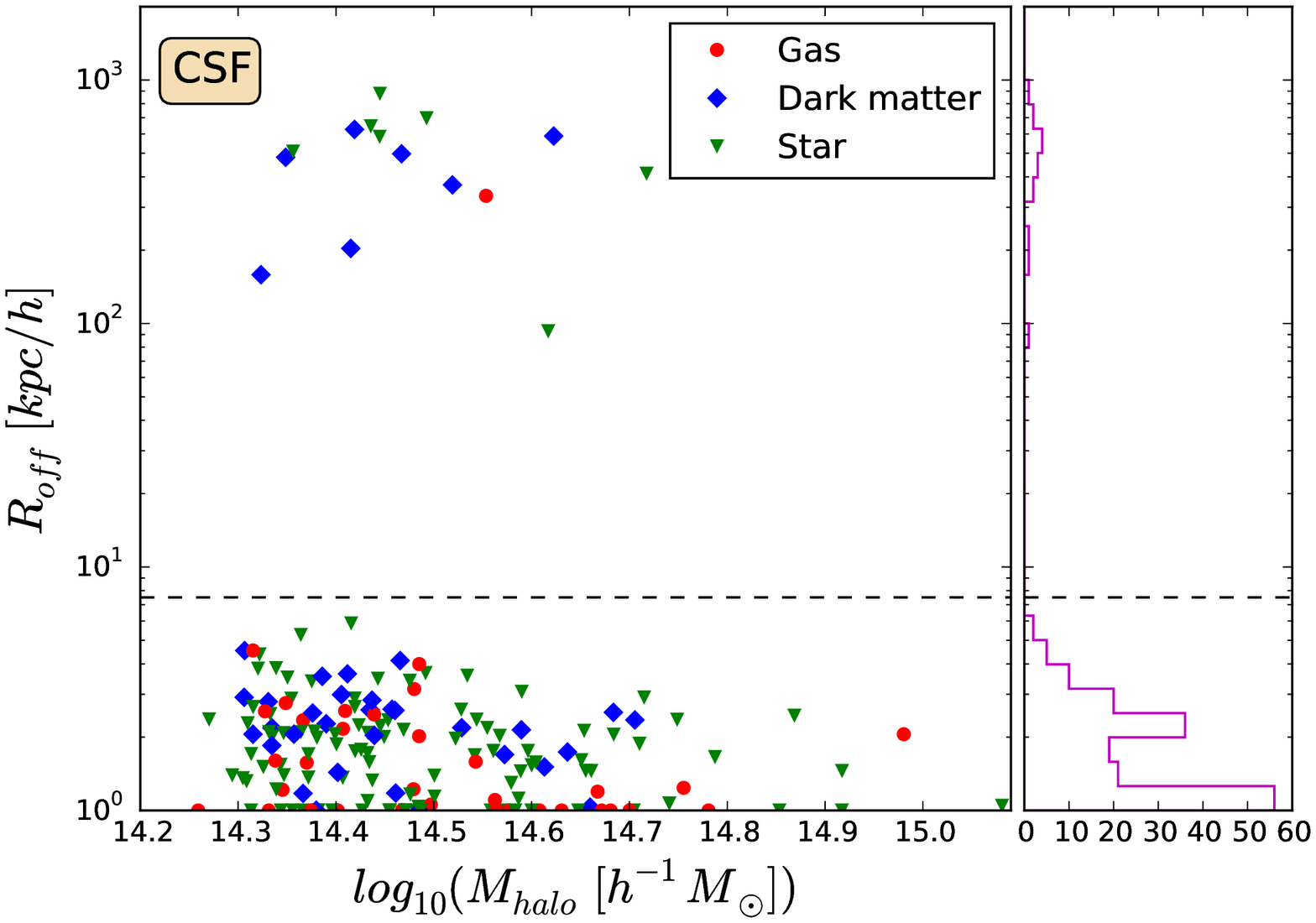} \\
  \includegraphics[width=0.5\textwidth]{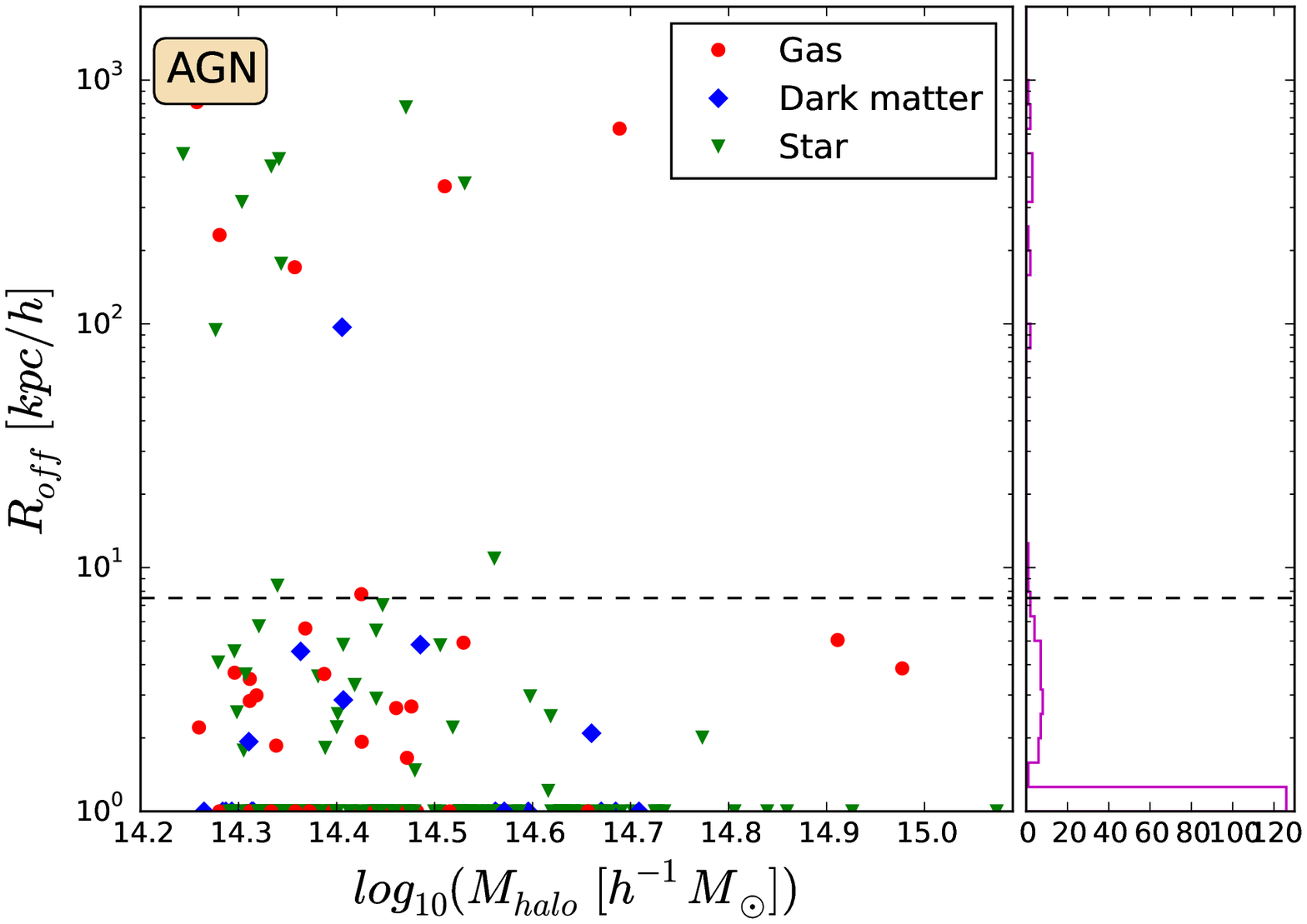}
\end{tabular}
\caption{The offset between the maximum SPH density and minimum potential positions as a function
  of halo mass. From top to bottom, these panels are for DM, CSF, AGN runs,
  respectively; On the right hand of each panel, we show a histogram distribution
  of the offsets. The horizontal dashed lines are the softening length in the
  simulations. As indicated in the legends of middle and bottom panels, the different
  color symbols represent the type of the highest density particle, i.e. cluster centre.}
  %, with dotted lines for unrelaxed clusters and solid lines for
  %relaxed clusters.}
\label{fig:Odp}
\end{figure}

\subsection{Offsets between maximum SPH density and minimum potential positions}

%%%%%%%%%%%%%%%%%%%%%%%%%%%
\begin{table}
\begin{centering}
\begin{tabular}{l||c||c||c|}
\hline
 & Gas & Dark matter & Star\\
\hline
\hline
  CSF & 36 & 38 & 110 \\
\hline
  AGN & 50 & 18 & 116 \\
\hline
\end{tabular}
\caption{Numbers of clusters in which the densest particle belongs to a given
  particle type (i.e. gas, dark matter or stars).}
\label{tabledpt}
\end{centering}
\end{table}
%%%%%%%%%%%%%%%%%%%%%%%%%%%

In Fig. \ref{fig:Odp}, we investigate the offset between the maximum of SPH
density and the minimum of the gravitational potential positions in the DM, CSF
and AGN runs (upper, middle and lower panels respectively). We reset offsets
$R_{off} < 1 \Kpc$ to $1 \Kpc$, for an easier visualization.

\begin{itemize}
\item In the DM run, we find typical offsets of $\sim 10 \Kpc$, which is
  comparable with
  the simulation softening length as indicated by the horizontal dashed line in all
  panels. Those clusters with large offsets contain massive compact substructures that
  are in the process of merging and the system shows obvious signs of disturbance.

\item In the CSF run, the typical offsets are smaller than the softening length
  of the simulation ($\simlt 3 \Kpc$), but in some cases there are offsets as large
  as $\simgt 100 \Kpc$. Close inspection shows that star and dark matter
  particles tend to be the particles defining the maximum SPH density within these
  systems; we indicate this explicitly by marking the particles that trace the
  maximum of the density with symbols defined in the legend.

\item In common with the CSF run, the majority of clusters in the AGN run have
  offsets smaller than the softening length,
  $\simlt 1 \Kpc$. As in the CSF run, and as shown in table \ref{tabledpt}, star
  particles tend to define the location of the density peak.

\end{itemize}

\noindent We have visually inspected those clusters that have large offsets in
Fig. \ref{fig:Odp} and find, unsurprisingly, that the density peak is associated
with a massive satellite galaxy (e.g. the disturbed cluster in the upper row of
Fig. \ref{fig:forshow}). This indicates that these clusters with large offsets
are normally undergoing major mergers and are visually disturbed.

%%%%%%%%%% FIG 3 %%%%%%%%
\begin{figure*}
  \centering
  \begin{tabular}{@{}cc@{}}
    \includegraphics[width=0.5\textwidth]{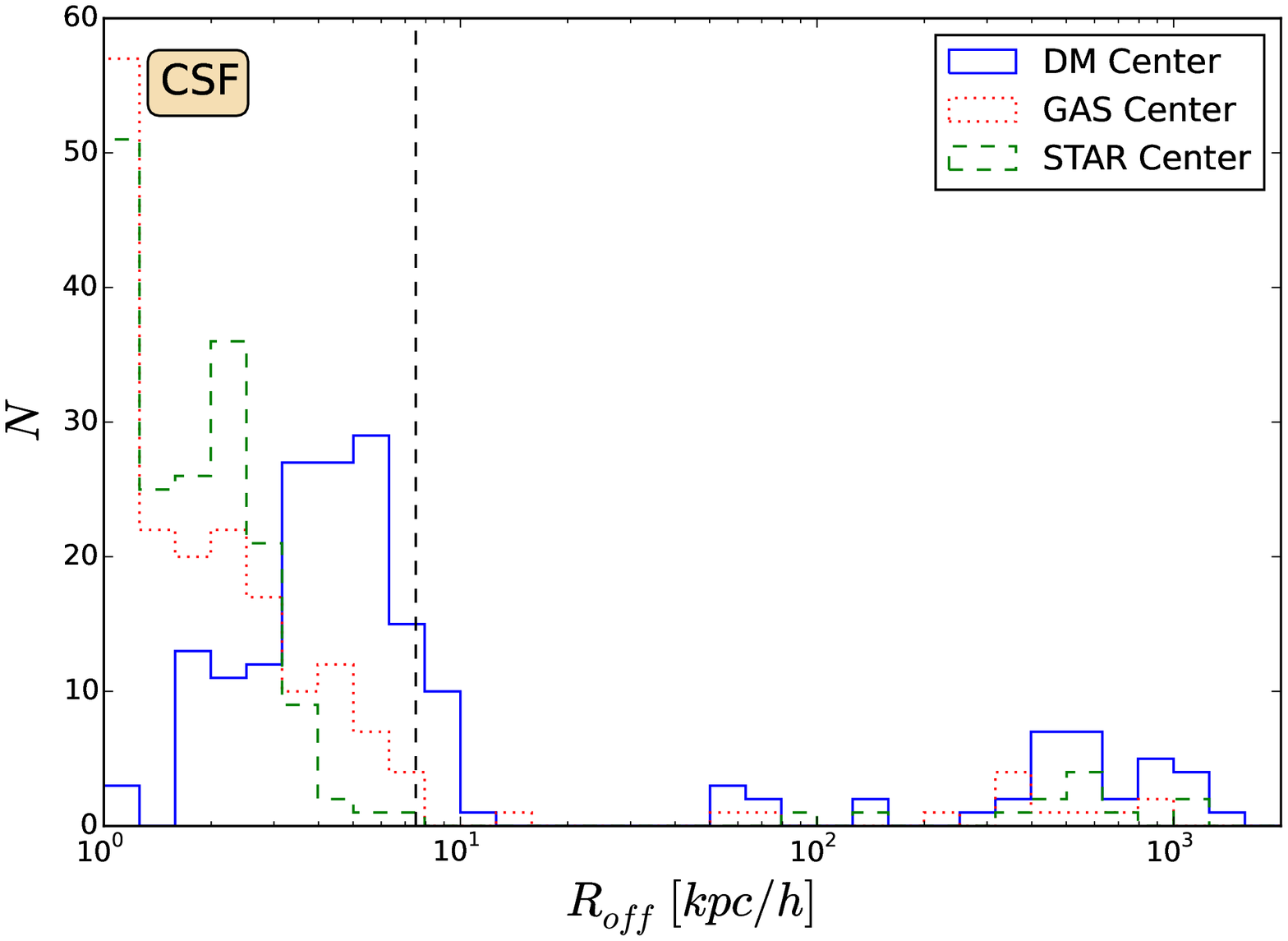}
    \includegraphics[width=0.5\textwidth]{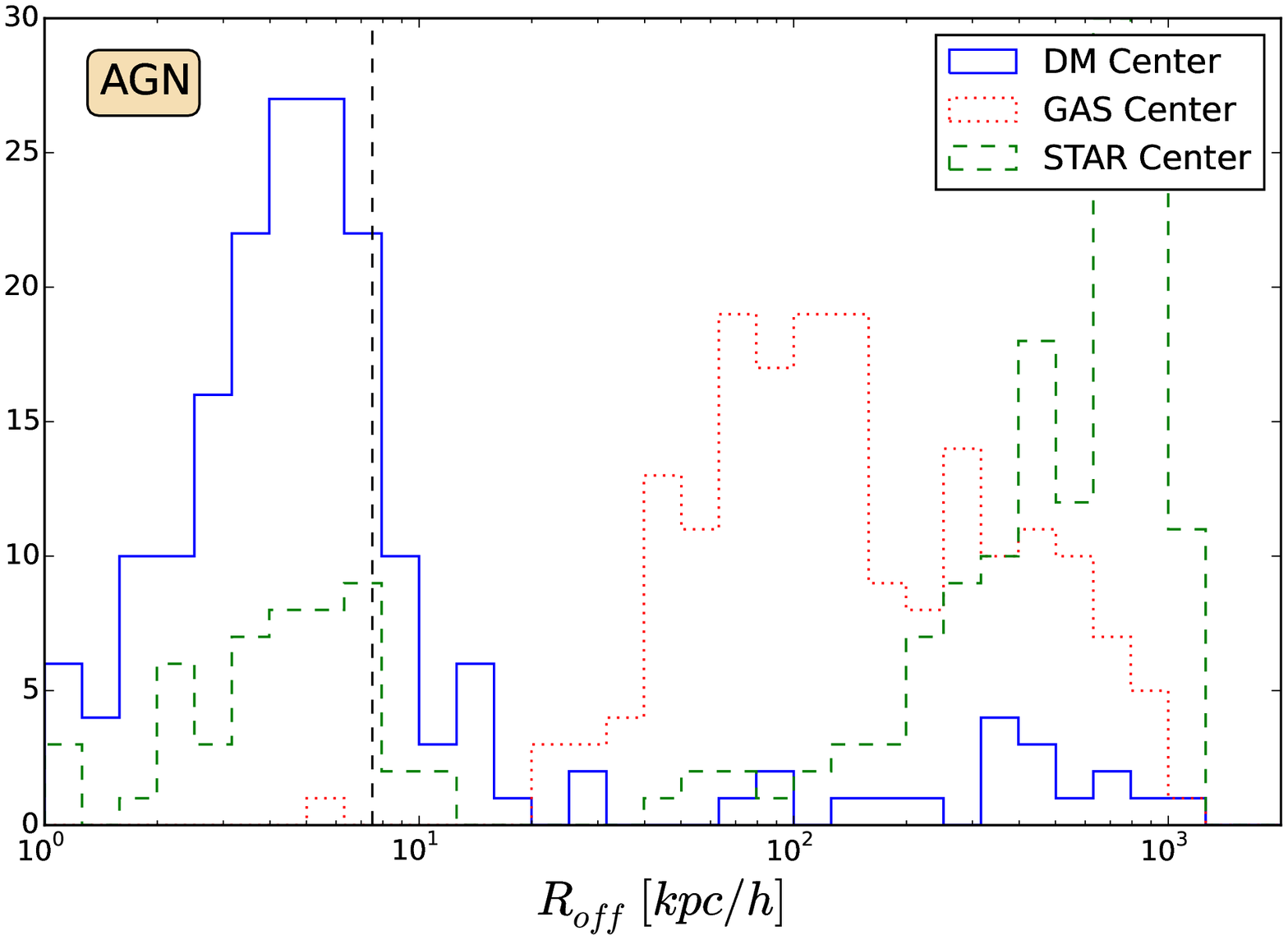}
  \end{tabular}
  \caption{As in Fig. \ref{fig:Odp}, we show the offset between the density and
    potential centres in the CSF (left panel) and AGN (right panel) runs, but now
    we split according to particle type, where solid, dotted and dashed histograms
    correspond to dark matter, gas and stars respectively. Note that
    offsets $R_{off} < 1 \Kpc$ are reset to $1 \Kpc$.}
  \label{fig:ddp}
\end{figure*}
%%%%%%%%%%%%%%%%%%%%%%%%%

We did not differentiate between the material that contributes to the
estimate of the maximum SPH density position (i.e. gas, star and dark matter particles are
given equal weight) in Fig. \ref{fig:Odp}; we now show this in Fig. \ref{fig:ddp}.
Here the maximum SPH density positions computed from each of the three particle types are
offset with respect to the potential centre of the cluster in the CSF and AGN
runs (left and right panels respectively). In this calculation, we
include only particles of the same species (i.e. dark matter, gas, stars)
when calculating densities. The particle with the maximum SPH density
is selected as the density peak for the given component.

In the CSF run, there is broad agreement between the maximum SPH density and minimum potential
position offsets computed for each of the particle types; these offsets are within
$\sim 10 \Kpc$, while those systems with offsets
$\gtrsim 300 \Kpc$ are visually idenitified as disturbed. In stark contrast to
the CSF run and also to the result from Fig. \ref{fig:Odp}, in the AGN
run there is a clear separation in the maximum SPH density and minimu potential position offsets
computed from dark matter particles on the one hand and star and gas particles on
the other. The dark matter particles have offsets similar to those found in the
CSF run, clustering within $\sim 10\Kpc$, but the star particles have two
offset peaks at $\sim 7$ and $\sim 900 \Kpc$, while gas particles
particles have offsets spread between $\sim 50 - 800 \Kpc$.
The large offsets we see in the stellar component arise because the
identified centres are located in satellite galaxies, which are compact,
rather than in the BCG. This is also linked to the large offsets we find
in the gas component, which arise because strong AGN feedback can expel
gas to a large cluster-centric radius and helps to suppress star formation
over much of the lifetime of the BCG by inhibiting the accumulation of
dense gas at small radii. Similar trends arising from AGN have been
reported in \cite{Cin2012, Cinthia13, Cui2014a}. Note that
this figure is primarily of theoretical interest; it shows how the centre of
density changes as we sample the different components in the simulation,
something that would be challenging to do observationally!

\subsection{Offsets between BCG and potential centres}

%%%%%%%%%% FIG 4 %%%%%%%%
\begin{figure*}
\centering
\begin{tabular}{@{}c@{}}
  \includegraphics[width=0.5\textwidth]{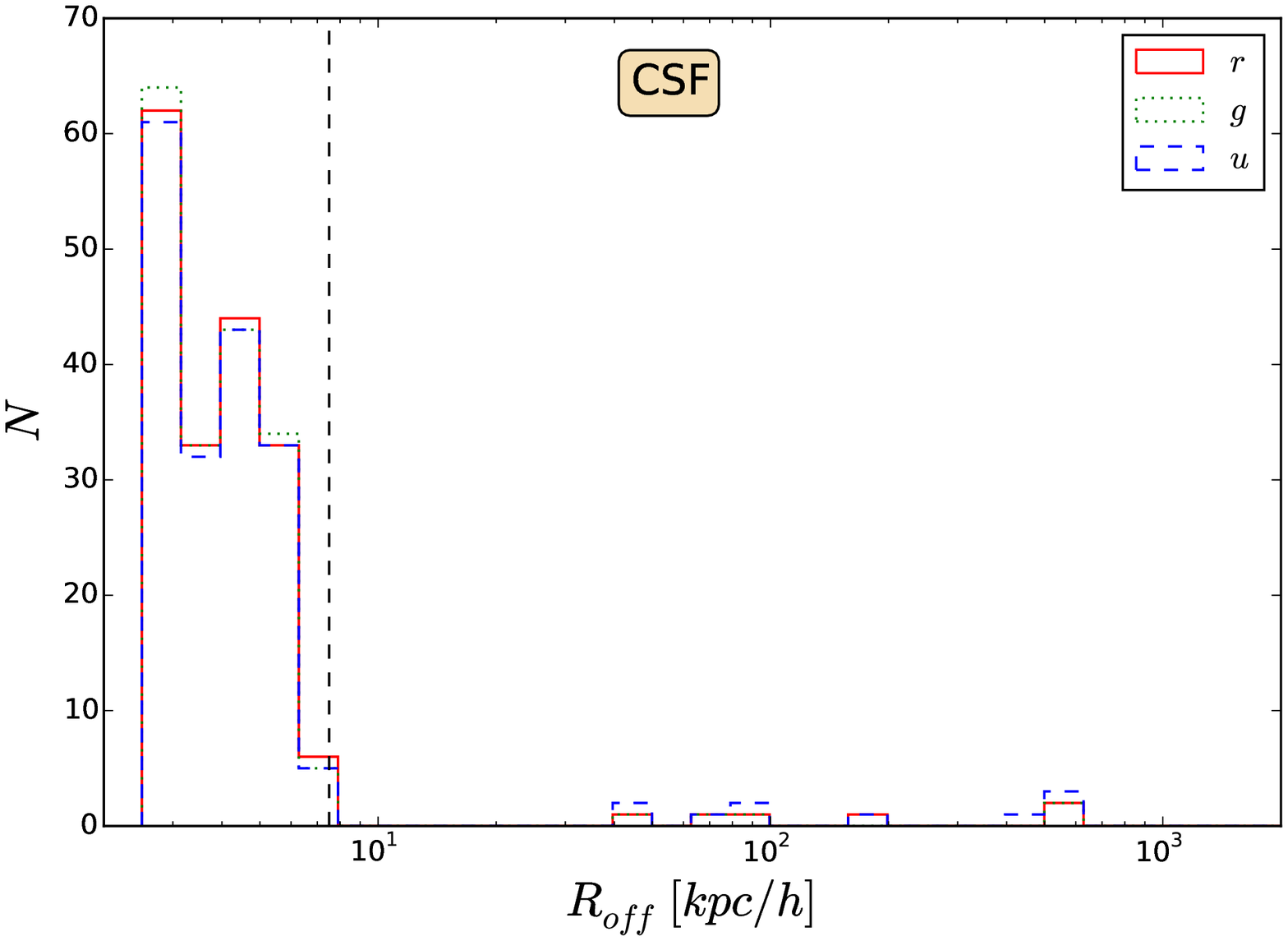}
  \includegraphics[width=0.5\textwidth]{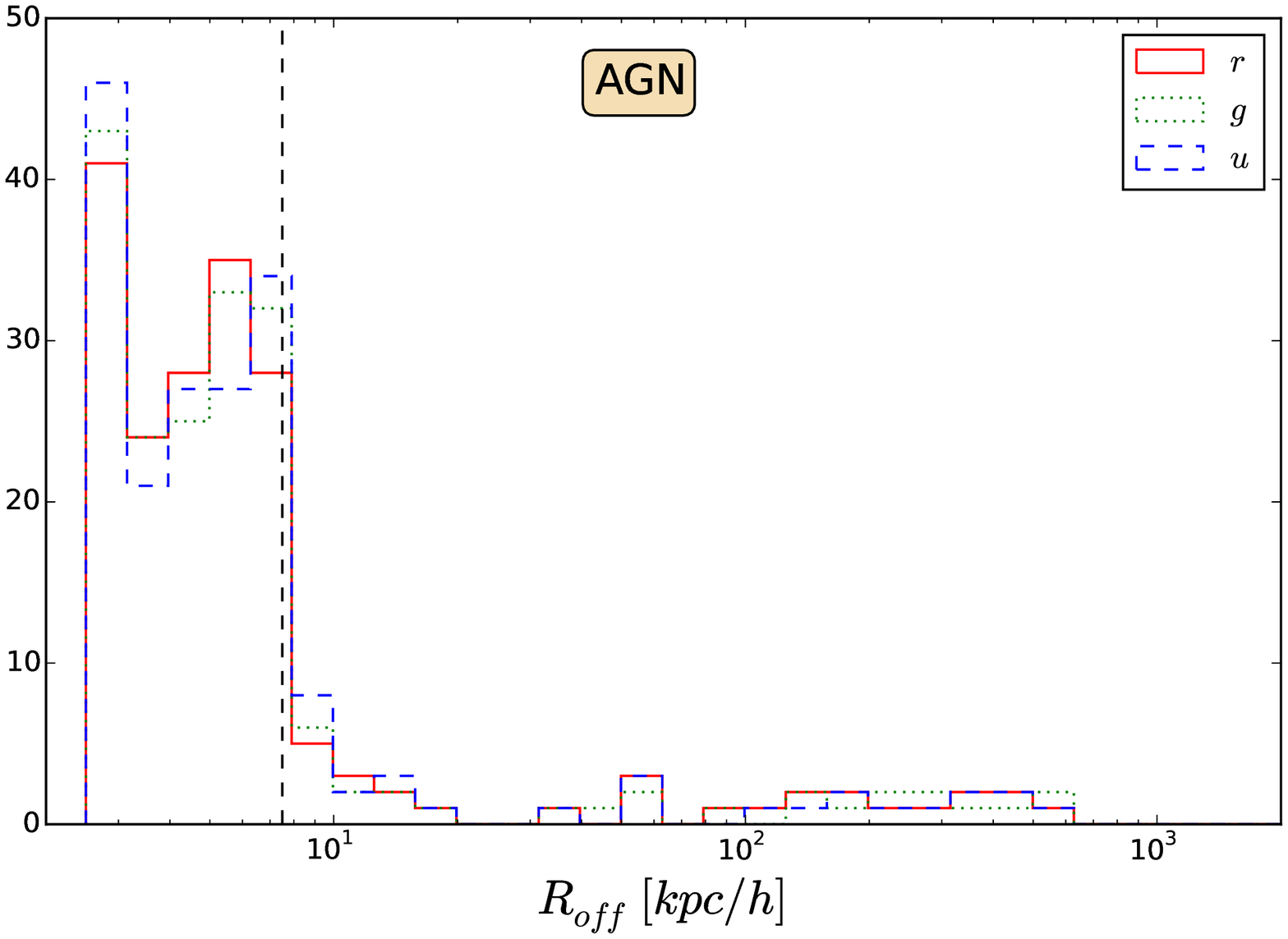} \\
\end{tabular}
\caption{The histogram of the offsets between centre of BCGs and
  cluster potential centre. Left panel is the results from CSF clusters, while
  the right panel is for AGN clusters. Three optical luminosity bands $u$, $g$,
  $r$ are indicated in the upper-right legend. The vertical dashed lines are the
  softening length in the simulations.}
\label{fig:dlp}

%%Fig 5
\begin{tabular}{@{}c@{}}
  \includegraphics[width=0.5\textwidth]{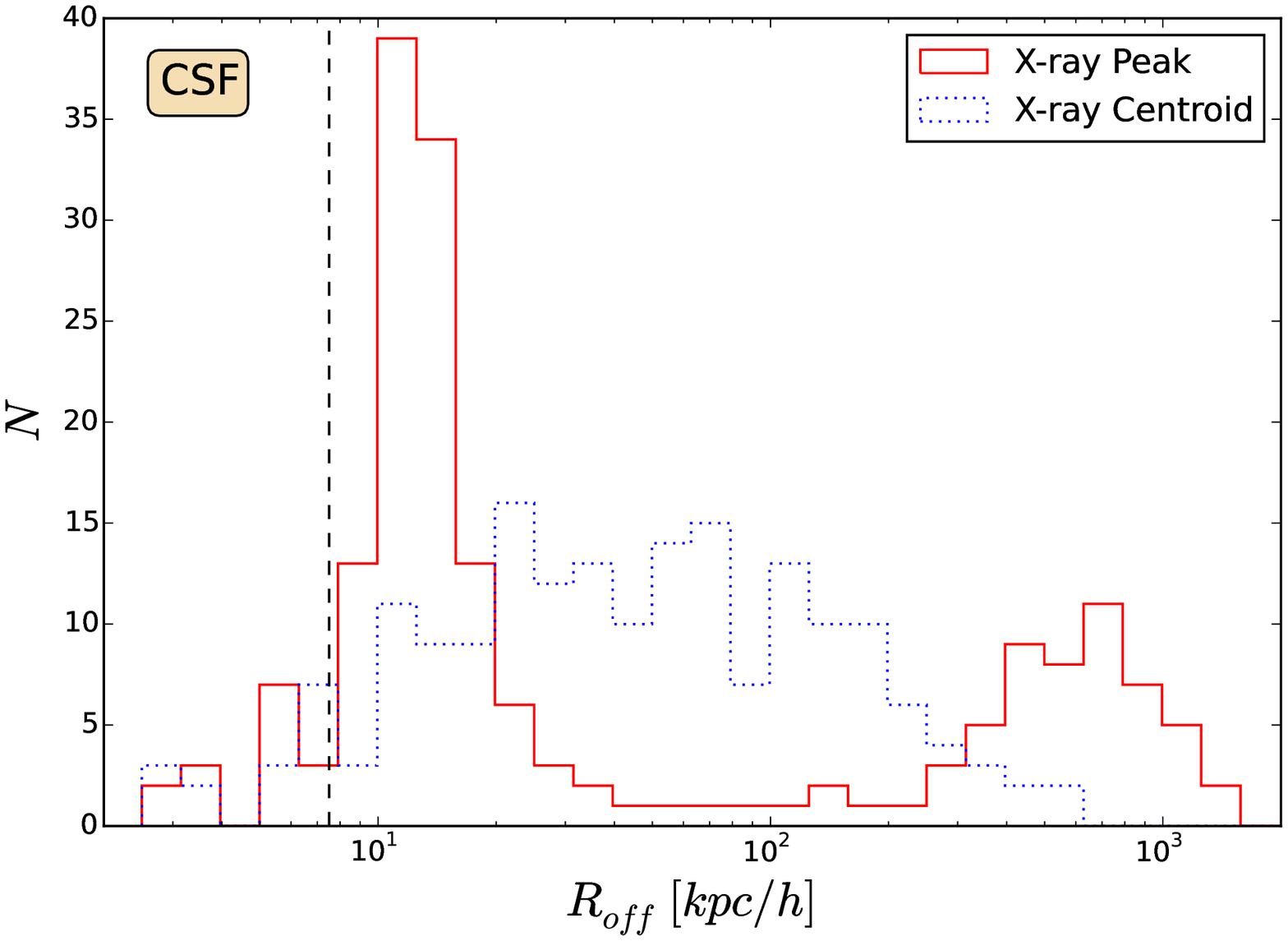}
  \includegraphics[width=0.5\textwidth]{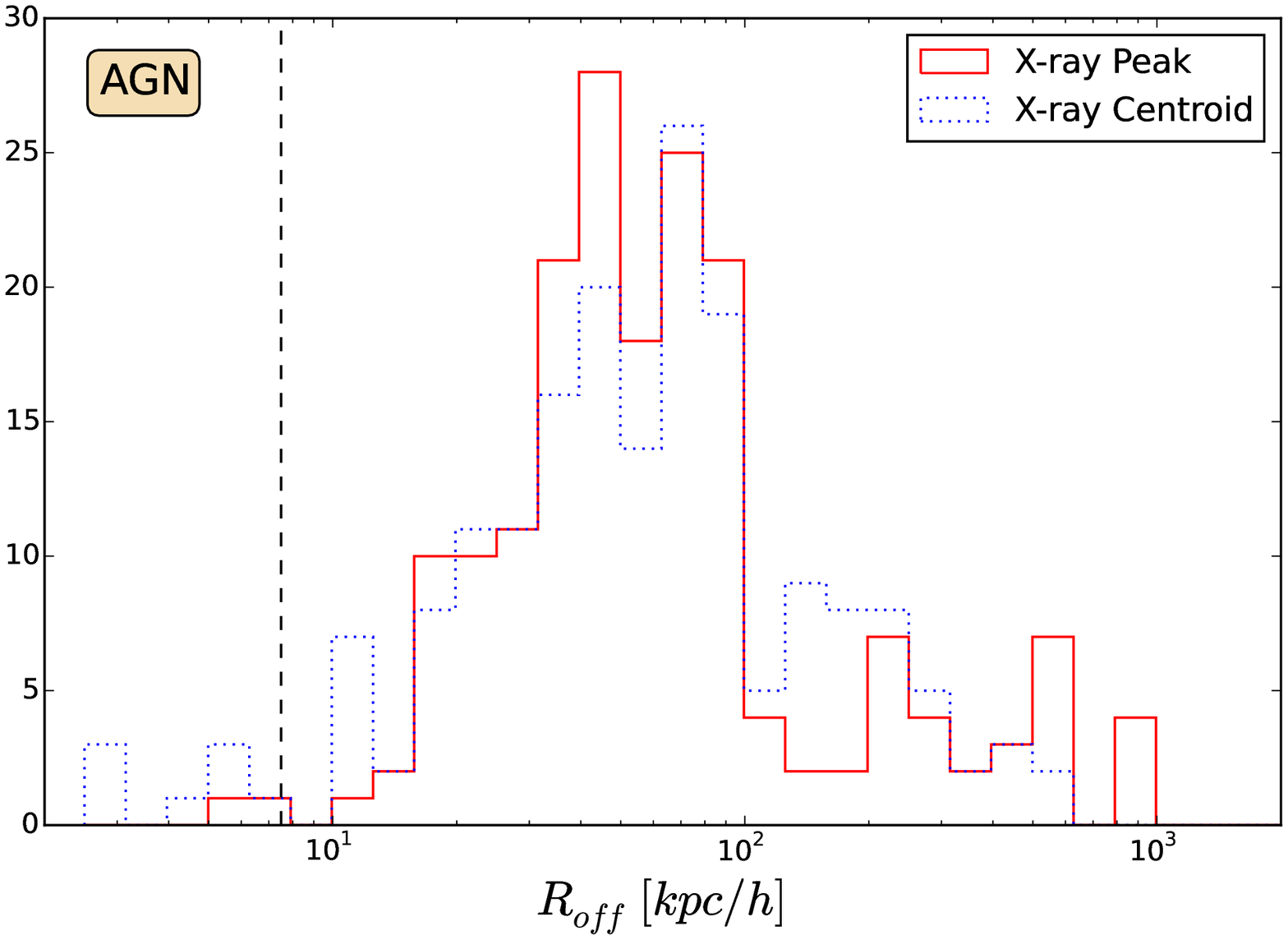}
\end{tabular}
\caption{Similar to Fig. \ref{fig:dlp}, the histogram of the offsets between X-ray centres
  and the cluster potential centre. Left panel is the results from CSF clusters, while the
  right panel is for AGN clusters. The peak and centroid indicators are shown in the
  legend. The vertical dashed lines are the softening length in the simulations.}
\label{fig:dxp}
\end{figure*}
%%%%%%%%%%%%%%%%%%%%%%%%%

We now consider the relationship between the centre of BCGs and minimum potential
positions, where we employ the method of \cite{Cui2011} as described in \S~\ref{method} to
assign luminosity to star particles in the SDSS $u$, $g$ and $r$ bands. Note that we do
not include the effects of dust when calculating luminosities, and so we potentially omit
band-dependent dust attenuation that could, in principle, bias our
conclusion. To compare with observations, we focus on 2D $x$-$y$ projections
here. The minimum offset is set to half of the pixel size $2.5 \Kpc$.

In Fig. \ref{fig:dlp} we show how the distribution of offsets between the
centre of BCGs and minimum potential positions. The results for both
the CSF and AGN runs (left and right panels respectively) are in broad
agreement, and similar to those shown in Fig. \ref{fig:Odp} for the offset
between density and potential peaks; most of the offsets are within the
softening length for both CSF and AGN runs. We find
no dependence on measured (i.e. $u$, $g$ or $r$) band.

\subsection{Offsets between X-ray and potential centres}

In Fig. \ref{fig:dxp}, we show the distribution of offsets between the X-ray peak,
centroid positions and cluster potential centre in the CSF and AGN runs (left and right
panels respectively). Here we note some interesting differences.

\begin{itemize}
\item In the CSF run, the offset distributions of peak positions show a
  peak at $\sim 10 \Kpc$, with a second peak at $\sim 700 \Kpc$; this is larger
  than for the offset between the centre of BCGs and minimum potential positions. While,
  the X-ray centroid offset show a wide spread distribution from $\sim 10 \Kpc$
  towards $\sim 200 \Kpc$.

\item In the AGN run, the offset distributions for both X-ray peak and centroid
  have a peak at $\sim 50 \Kpc$. This is slightly less than the offsets between
  the gas component density and cluster potential centres from
  Fig. \ref{fig:ddp}.
  %% The centroid offsets distribution is similar between the AGN and CSF runs.
  %% This difference in the
  %% offset distribution between the CSF and AGN clusters reveals the result of the powerful
  %% AGN feedback, which removes hot gas -- the origin of the X-ray emission -- from the
  %% central regions and deposits it at larger radii.
  %% Also, the large offsets that
  %% characterize some of the unrelaxed clusters is often due to a peak in the photon map
  %% that is located in a sub-halo of the main system.

\end{itemize}

\noindent Compared with the X-ray peak centre, the centroid is more stable for both
CSF and AGN runs. They tend to have similar distributions, despite the AGN feedback
model. However, the centroid offsets from CSF runs have no clear peak compared
to the AGN runs. There is no strong evidence of the secondary peak for the
centroid offsets. The X-ray peak offsets for CSF run are smaller than the AGN
run, which indicates that the AGN feedback has stronger effects on the X-ray
peak position.\\

\noindent These results suggest that the centre of BCGs should be a more reliable and
precise tracer of the underlying gravitational potential, and is also less likely to be
influenced by the AGN feedback.

\section{Discussion and Conclusions}
\label{concl}

Using a suite of cosmological $N$-body and hydrodynamical simulations, we have constructed
a mass- and volume-complete simulated galaxy cluster catalogue. We have considered
a pure dark matter (i.e. $N$-body only) model and two galaxy formation models that include
cooling, star formation and supernova feedback, with and without AGN feedback (the CSF and
AGN runs respectively); this allows us to explore in a systematic fashion the impact of
these two baryon models on the properties of galaxy clusters. In this paper, we have
assessed how estimates of galaxy cluster centres are influenced by the mode of
measurement -- using X-ray emitting hot gas, the centre of BCGs, or
the total mass distribution, which is accessible via gravitational lensing,
say. In all cases we compare to the location of the minimum of the gravitational
potential of the system, which we would expect to define a physically reasonable
centre of the system, assuming that it is in dynamical equilibrium.

The main results of our analysis are summarised as follows.
\begin{itemize}

\item We find that the maximum local density, computed using an SPH kernel
  smoothing over 128 nearest neighbours, is in good agreement with the minimum
  of the gravitational potential regardless of the assumed galaxy formation
  model, \emph{provided we include all particles -- dark matter, gas and stars
    -- in the calculation}. In the CSF runs, we find offsets between the maximum SPH density
  and minimum potential positions at $\simlt 3 \Kpc$; in the AGN runs, these offsets are
  even smaller than the CSF runs. However, both runs have a small amount of
  clusters with very large offsets ($\simgt 100 \Kpc$). This is because the
  density peak is associated with a satellite galaxy.

  If we compute the maximum local density for individual particle types, we find
  differences that depend on the assumed galaxy formation model. The offsets for
  different particles in CSF run are within the simulation softening
  length. However, many clusters in AGN run have very large offsets between the
  density peak evaluated from both stellar and gas particles and the potential
  centre. The strong feedback from the AGN not only expels gas particles, which
  have the offset at $\sim 100 \Kpc$, but also reduces the stellar density
  within the central galaxy, in which case the peak of the stellar density is
  more likely to be associated with a satellite galaxy.

\item Using projected optical luminosities in SDSS $r$, $g$ and $u$ bands, we
  identify the centre of BCG from star particles in the CSF and AGN runs. We
  find that centre of BCGs are close to the potential centre, within the softening length in both
  runs and independent of the assumed band. A small fraction of the clusters
  have large offsets in both CSF and AGN runs; these belong to disturbed
  clusters, in which the identified BCG is offset from the centre of the
  potential by visually checking.

\item Identifying the location of both the peak and centroid positions of X-ray
  emission from realistic maps, we find slightly larger peak offsets $\sim 10
  \Kpc$ in the CSF run (with a second peak at $\sim 700 \Kpc$); $\sim 50 \Kpc$
  in the AGN run. The X-ray centroid offset seem more stable than X-ray peak,
  which have less effect from the AGN feedback. It has a wide spread from $\sim
  10 \Kpc$ to $\sim 200 \Kpc$. There is no clear peak in the CSF run; while the
  AGN run has a similar peak as its X-ray peak offset.

  %% X-ray centroid offsets seem to have a wider spread than
  %% X-ray peak offsets. While the latter has a second peak at $\sim 800 \Kpc$ in
  %% the CSF run. Offsets from the X-ray peak and centroid position have similar
  %% peak values: $\sim 50 \Kpc$ in the AGN run. We interpret this larger offset as
  %% a natural consequence of powerful AGN feedback displacing dense gas from the
  %% central regions of the cluster.

\end{itemize}

It is interesting to ask how well our simulations match observations,
  which has a bearing on the general applicability of our results. We note that
  we have already used the same cosmological simulation data to compare
  baryon and stellar mass fractions with observations in \cite{Cui2014b} (see
  their Fig. B1 for details). There it was shown that both of these fractions
  computed from our AGN simulation are consistent with observations, whereas
  the CSF runs predict values that are larger than observed; this is to be
  expected, arising because of overcooling. In the nIFTy cluster comparison
  project \citealt{Sembolini2015a, Sembolini2015b},
  a single galaxy cluster has been simulated in a cosmological context with
  a range of state-of-the-art astrophysics codes, and in the runs that employ
  the physics of galaxy formation (e.g. radiative cooling, star formation,
  feedback from supernovae and AGN) it has been shown the results from
  the model used in this paper is consistent with the results of other codes
  (\citealt{Sembolini2015b}; Cui et al., In Preparation) in global cluster properties.
  However, galaxies inside this
cluster show striking code-to-code variations in stellar and gas mass
\citep{Elahi2015}, which implies different spatial
distributions for the gas and stellar components. Thus, we caution that
the choice of input physics in simulations of this kind can have a strong
quantitative influence on the results.
  %% It is worth
  %% noting, however, that a generic issue with galaxy formation models that
  %% employ strong AGN feedback is that they produce BCGs with too large
  %% effective radii \citep{Cinthia13} and too low central gas densities
  %% (Cui et al. 2015, In Preparation).}

We find that the distribution of offsets between the centre of BCG and
X-ray emission centres with respect to the potential centre is smaller than is
found observationally; this could be due to in part to observational
inaccuracies (image resolution, identification of lensing centre) and in part to
our assumption that the potential centre, calculated from the 3D distribution of
matter within the cluster, is well-matched to the lensing centre. However, our
results agree with observations that centre of BCG is a better
tracer of the cluster centre than the X-ray emission weighted centre
\citep[][]{George2012}.  However, the claim that the BCG is a better
  tracer requires identifying BCGs correctly in the first place in
  observations, which is not straightforward. Our simulation results suggest
  that the simple grouping method after ICL extraction in Section \ref{method}
  does a good job.
Offsets between X-ray and lensing centres are in fact
observed at a level of 100 kpc \citep[e.g.][]{Allen1998, Shan2010a,
  George2012}. However, the observed offsets between lensing and BCGs are
usually smaller. For example, \cite{Oguri2010} found that the offsets between
weak lensing and BCG are at $\sim 50 \Kpc$, while the strong lensing has even
closer position to BCG \citep{Oguri2009}. With large statistical samples,
\cite{Zitrin2012} also suggested smaller offsets between the weak lensing and
BCG position. These support that the BCG traces the minimum gravitational
potential position better than the X-ray data.

The large offset tail found in clusters from both the centre of BCG and
X-ray center are basically consistent with the secondary peak found by
\cite{Johnston2007, Zitrin2012}. These large offsets should be caused by
dynamically unrelaxed clusters undergoing mergers, in which the optical luminosity and
X-ray centres can be located at a massive satellite galaxy, which is away from
the cluster potential centre. Using a set of hydrodynamical simulations of
mergers of two galaxy clusters, \cite{Zhang2014} find that significantly large
SZ-X-ray peak offsets (\textgreater 100 kpc) can be produced during the major
mergers of galaxy clusters. This finding is basically agreed to the second peak
for X-ray peak-potential offsets from our CSF runs. These large offsets indicate
these clusters are not relaxed. This highlights the
importance of dynamical state in the centre determination,
something we will address in a follow-up paper.

Finally, we have considered only spatial offsets in this study, the first of a
series. We expect to find dynamical offsets within clusters. Subhaloes or
satellite galaxies in N-body and hydrodynamic simulations are found to have
velocities differing from the dark matter halos \citep[e.g.][]{Diemand2004,
  Gao2004, Gill2004b, Munari2013, Wu2013}. These velocity offsets are closely
connected to the cluster center offsets. \cite{Gao2006, Behroozi2013}
demonstrated that dark matter halo cores are not at rest relative to the halo
bulk or substructure average velocities and have coherent velocity offsets
across a wide range of halo masses and redshifts. We revisit this using our
cluster sample in our next paper, surveying not only the dark matter but also
gas and stars, and consider its implications for turbulence and accretion onto
AGN.

\section*{Acknowledgements}

We thank the referee for their thorough and thoughtful review of our paper.
All the figures in this paper are plotted using the python
matplotlib package \citep{Hunter:2007}. Simulations have been carried
out at the CINECA supercomputing Centre in Bologna, with CPU time
assigned through ISCRA proposals and through an agreement with the
University of Trieste.  WC acknowledges the supports from University of
Western Australia Research Collaboration Awards
PG12105017, PG12105026, from the Survey Simulation Pipeline (SSimPL;
{\texttt{http://www.ssimpl.org/}}) and from iVEC's Magnus supercomputer under
National Computational Merit Allocation Scheme (NCMAS) project gc6.
WC, CP, AK, GFL, and GP acknowledge support of ARC DP130100117.
CP, AK, and GFL acknowledge support of ARC DP140100198.
CP acknowledges support of ARC FT130100041.
VB, SB and GM acknowledge support from the PRIN-INAF12 grant ’The Universe in a
Box: Multi-scale Simulations of Cosmic Structures’, the PRIN- MIUR 01278X4FL
grant ’Evolution of Cosmic Baryons’, the INDARK INFN grant and ’Consorzio per la
Fisica di Trieste’. AK is supported by the {\it Ministerio de Econom\'ia y
  Competitividad} (MINECO) in Spain through grant AYA2012-31101 as well as the
Consolider-Ingenio 2010 Programme of the {\it Spanish Ministerio de Ciencia e
  Innovaci\'on} (MICINN) under grant MultiDark CSD2009-00064.  He further thanks
The Cure for faith. GP acknowledges support from the ARC Laureate program of
Stuart Wyithe.

%*****************************************************************************
\bibliographystyle{mn2e.bst}
\bibliography{bibliography}

\appendix

\section{Identifying Density Peaks}
\label{A:denspeak}

%%%%%%%%%% FIG 4 %%%%%%%
\begin{figure}
\includegraphics[width=0.5\textwidth]{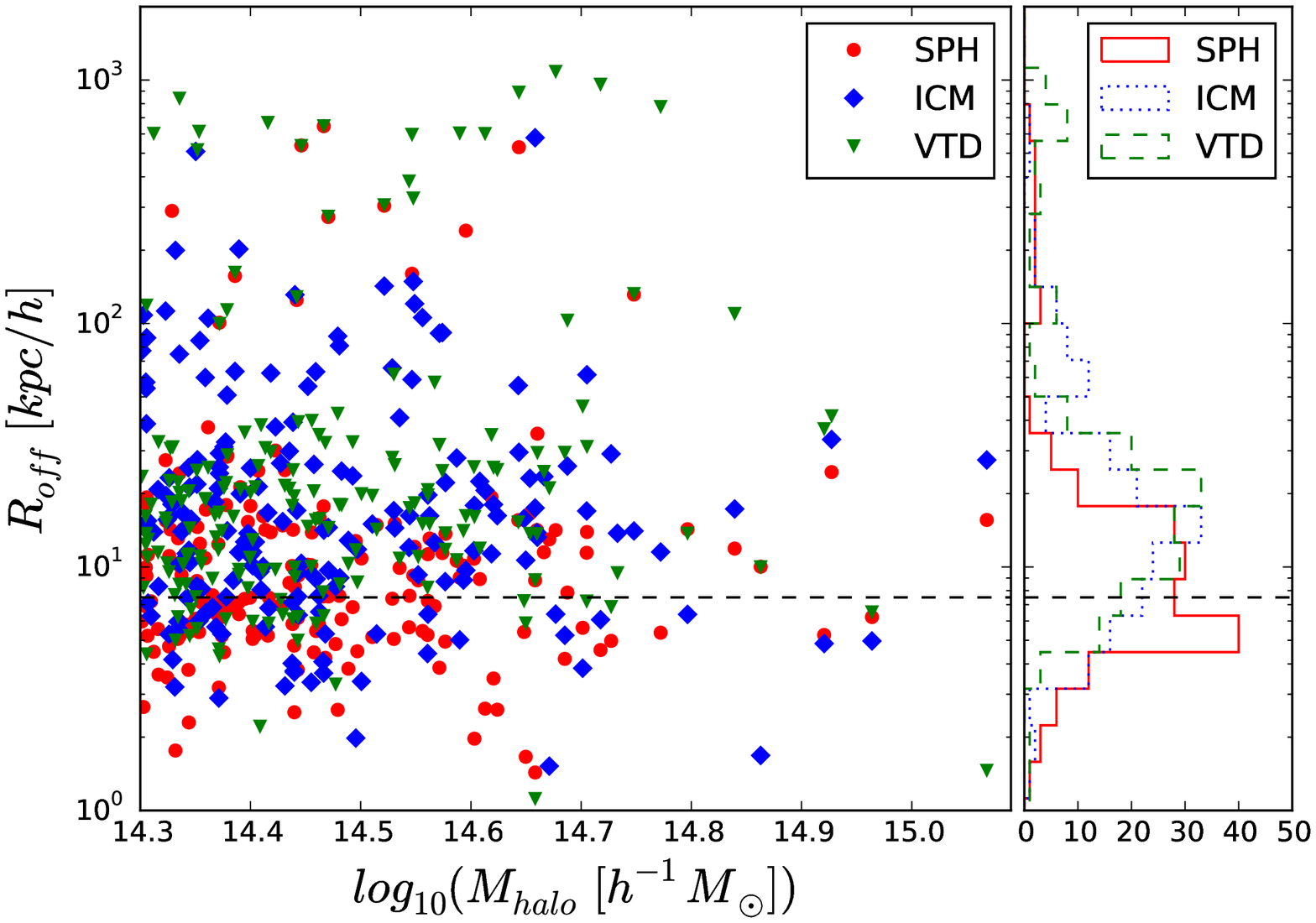}
\caption{The offsets between three density peak estimators -- SPH, Voronoi tessellation
  density (VTD), and the iterative centre of mass (ICM) -- and the location of the minimum
  of the gravitational potential, as a function of cluster mass. Red circles, blue
  diamonds and green inverted triangles correspond to SPH, ICM and VTD estimators
  respectively. The right panel shows the histograms of the offsets. See
  text for further details.}
\label{fig:a1}
\end{figure}
%%%%%%%%%%%%%%%%%%%%%%%%%

We considered a number of approaches to estimating the location of the maximum density
of the cluster. Here we briefly review three -- one that was used in the study, and two
others from the literature.

\begin{itemize}
\item The Smoothed Particle Hydrodynamics (SPH) method adopts the kernel smoothing
  approach that is commonly used in hydrodynamics; we have implemented and tested this
  method in \cite{Cui2014b} using 128 neighbours when calculating densities. This is the
  method used in the {\small{PIAO}} halo finder and the one used in this study.

\item The Iterative Centre of Mass (ICM) method estimates the mass-weighted centre in an
  iterative fashion, using all particles within a shrinking spherical volume until
  convergence in the estimated centre is achieved \citep[cf.][]{Power2003}; we define
  convergence when consecutive centres agree to within $1\Kpc$.

\item The Voronoi Tessellation Density (VTD) method partitions the volume into cells using
  the distance between adjacent points to define cell boundaries, and uses the inverse
  volume of the cell to estimate the local density at the position of each particle; it
  requires no free parameters. We use the publicly available convex hulls program
  \citep{Clarkson1992} implemented in {\small{python}}. We note that this approach is
  sensitive to the finite resolution of the simulation.
\end{itemize}

\noindent In Fig. \ref{fig:a1}, we show the offsets between the three estimates (SPH, ICM,
and VTD) of the maximum density position and the location of the minimum of the gravitational
potential (red circles, blue diamonds and green inverted triangles, respectively) for each
of the clusters in our DM sample. The histograms in the right hand panel are the
corresponding to projected distributions of cluster offsets. Fig.~\ref{fig:a1} shows that
the performance of the three estimators, as measured by the typical size of offset with
respect to the location of the minimum of the gravitational potential, is comparable,
although the SPH method -- implemented in {\small{PIAO}} and used in this study -- should
be favoured -- $87.5$ per cent of the total offsets are within $20 \Kpc$.

\bsp
\label{lastpage}
\end{document}